\documentclass[sigconf, 9pt, nonacm]{acmart}

\usepackage{threeparttable}
\usepackage{amsthm}
\usepackage{textcomp}
\usepackage{booktabs}
\usepackage{bm}
\usepackage{multirow}
\usepackage{algorithm}
\usepackage{epstopdf}
\usepackage{epsfig}
\usepackage{enumitem}
\usepackage{booktabs}
\usepackage{diagbox}
\usepackage{amsmath,amsthm}

\usepackage{multirow}
\usepackage[normalem]{ulem}
\usepackage{makecell}
\usepackage[pscoord]{eso-pic}
\usepackage{xcolor}
\usepackage{stfloats}
\usepackage[table]{xcolor}
\usepackage{multirow}
\usepackage[normalem]{ulem}
\usepackage{makecell}

\newcommand{\qdel}[1]{}

\newcommand{\Qdel}[1]{}

\newcommand{\tdel}[1]{}

\newcommand{\hdel}[1]{}

\newcommand{\tabincell}[2]{\begin{tabular}{@{}#1@{}}#2\end{tabular}}

\AtBeginDocument{%
  \providecommand\BibTeX{{%
    \normalfont B\kern-0.5em{\scshape i\kern-0.25em b}\kern-0.8em\TeX}}}

\setcopyright{none}
\renewcommand\footnotetextcopyrightpermission[1]{} 
\begin{document}

\title{\textit{TrojanLoC}: LLM-based Framework for RTL Trojan Localization}

\author[Weihua~Xiao, Zeng~Wang, Minghao~Shao, Raghu~Vamshi~Hemadri, Ozgur~Sinanoglu, Muhammad~Shafique, Johann~Knechtel, Siddharth~Garg, Ramesh~Karri]{%
Weihua~Xiao$^\dagger$$^\S$,
Zeng~Wang$^\dagger$$^\S$,
Minghao~Shao$^\dagger$$^\ddagger$,
Raghu~Vamshi~Hemadri$^\dagger$,
Ozgur~Sinanoglu$^\ddagger$,
Muhammad~Shafique$^\ddagger$,
Johann~Knechtel$^\ddagger$, 
Siddharth~Garg$^\dagger$, 
Ramesh~Karri$^\dagger$\\[4pt]
$^\dagger$NYU Tandon School of Engineering, New York, USA\\
$^\ddagger$NYU Abu Dhabi, Abu Dhabi, UAE\\[3pt]
\text{\{wx2356, zw3464, shao.minghao, rh3884, ozgursin, muhammad.shafique, johann\}@nyu.edu,}
\text{siddharth.j.garg@gmail.com, rkarri@nyu.edu}
}


\begin{abstract}
\textit{Hardware Trojan}s (\textit{HT}s) are a persistent threat to integrated circuits, especially when inserted at the \textit{register-transfer level} (\textit{RTL}). 
Existing methods typically first convert the design into a graph, such as a gate-level netlist or an RTL-derived \textit{dataflow graph} (DFG), and then use a \textit{graph neural network} (\textit{GNN}) to obtain an embedding of that graph, which (i) loses compact RTL semantics, (ii) relies on shallow GNNs with limited receptive field, and (iii) is largely restricted to coarse, module-level binary HT detection.
We propose \textit{\textbf{TrojanLoC}}, an \textit{LLM-based framework for RTL-level HT localization}. 
We use an RTL-finetuned LLM to derive \textit{module-level} and \textit{line-level embedding}s directly from RTL code, capturing both global design context and local semantics. 
Next, we train task-specific classifiers on these embeddings to perform \textit{module-level Trojan detection}, \textit{type prediction}, and fine-grained \textit{line-level localization}. 
We also introduce \textit{TrojanInS}, a large synthetic dataset of RTL designs with systematically injected Trojans from four effect-based categories, each accompanied by precise line-level annotations. Our experiments show that \textit{\textbf{TrojanLoC}} achieves strong module-level performance, reaching 0.99 F$1$-score for Trojan detection, up to 0.68 higher than baseline, and 0.84 macro-F$1$ for Trojan-type classification. At the line level, TrojanLoc further achieves up to 0.93 macro-F$1$, enabling fine-grained localization of Trojan-relevant RTL lines.
\end{abstract}

\maketitle
\begingroup
\renewcommand\thefootnote{$\S$}%
\footnotetext{Authors contributed equally to this research.}%
\endgroup
\section{Introduction}
\label{sec:intro}

\textit{Machine learning} (\textit{ML}) has been increasingly applied to hardware security challenges~\cite{wang2025vericontaminated, wang2025verileaky, wang2025salad}, evolving from early classical ML methods~\cite{zareen2018detecting, huang2020survey} to graph-based techniques that leverage \textit{Graph Neural Network}s
(\textit{GNN}s) for modeling circuit structures~\cite{thorat2025Trojan, zhang2025gnn, chen2024gnn4ht}.
While these approaches have advanced \textit{hardware Trojan} (\textit{HT}) detection, the
emergence of \textit{large language model}s (\textit{LLM}s) introduces a new paradigm~\cite{wang2024llms, shao2024survey}.
LLMs provide semantically rich representations of hardware description languages for HT analysis.

Most recent learning-based defenses follow a common pipeline:
they first encode a hardware design into one or more vector
\emph{embeddings}, and then train a \textit{classifier} on these embeddings to decide
whether the design is Trojaned~\cite{Yasaei2021,Lashen23,zhang2025gnn,thorat2025Trojan}.
Within this pipeline, the classifier is
a separate component that operates on fixed-dimensional feature vectors, and the
core technical challenge is how to construct embeddings that capture
comprehensive RTL semantics relevant for downstream HT detection.
Existing methods typically address this by transforming RTL designs into
intermediate graph representations and then using a GNN to extract graph-level
embeddings~\cite{Yasaei2021,thorat2025Trojan}.
Two main types of intermediate graphs are commonly used:
(i) gate-level netlists obtained after synthesis~\cite{Lashen23}, and
(ii) RTL-derived \textit{dataflow graph}s (\textit{DFG}s)~\cite{Yasaei2021,thorat2025Trojan}.

This graph-construction step has three key drawbacks.
First, mapping RTL to gate-level netlists or DFGs discards RTL
semantics, e.g., a single RTL statement can expand into thousands of gates.
Second, GNNs are typically kept shallow to avoid over-smoothing, which limits
their receptive field and capture the global semantic of a RTL code.
Third, most existing methods focus on coarse, module-level binary detection and
offer little support for fine-grained localization, while gate-level approaches
also require full synthesis and incur significant overhead.
In parallel, recent works have shown that LLMs can serve as general-purpose
embedding extractors for input texts~\cite{Zhou24}, software codes~\cite{Bui25}, and RTL codes analysis~\cite{Hemadri25}.
Additionally, another type of works directly prompts LLMs with RTL code and asks them to judge whether a given design contains a Trojan, effectively using the LLM itself as a Trojan detector~\cite{Hayashi25}.

In this work, we use LLMs to extract embeddings directly from RTL codes, which
capture the design semantics needed for Trojan detection and localization.
To realize this LLM-based approach, we develop \emph{TrojanLoC: LLM-based Framework for RTL Trojan Localization}, which combines the TrojanInS dataset (Section~\ref{subsec:TrojanIns-Dataset}), LLM-based embedding extraction (Section~\ref{subsec:llm-embeddings}), and task-specific downstream classifiers (Section~\ref{subsec:classifier-training}) for module-level and line-level Trojan analysis. 
\textit{TrojanInS} is a synthetic RTL dataset with GPT-4.1 generated Trojans from four categories (T1–T4), providing module-level presence/type labels and line-level annotations after preprocessing (Section~\ref{subsec:TrojanIns-Dataset}).
TrojanLoC processes each RTL module at two levels. 
At the module level, the entire RTL text is encoded by an RTL-finetuned decoder-only LLM and mean-pooled a single module embedding (Section~\ref{subsubsec:module-embeddings}). 
At the line level, each RTL line is encoded to obtain a line embedding (Section~\ref{subsubsec:line-embeddings}). 
Then, both module-level and line-level embeddings are passed through two autoencoders for dimension reduction, which are finally used to train classifiers for both module-level and line-level Trojan tasks (Section~\ref{subsec:classifier-training}).
Overall, TrojanLoC uses LLM-derived RTL embeddings as a shared semantic backbone, while separate autoencoders and classifiers are trained on these embeddings for module-level detection, type classification, and fine-grained line-level localization.

Our main contributions are summarized as follows:
\begin{enumerate}[leftmargin=*]
  \item \textbf{TrojanInS Dataset.}
        We construct TrojanInS, a large-scale RTL dataset with 17k+ validated
        designs spanning four Trojan families (T1–T4), enabling comprehensive
        training and benchmarking of RTL-level Trojan detection and
        localization methods.

  \item \textbf{TrojanLoC Framework.}
        We propose TrojanLoC, a unified LLM+classifier framework that generates
        module-level and line-level RTL embeddings and, together with
        autoencoders and classifiers, supports module-level Trojan detection, Trojan-type prediction, and fine-grained line-level localization.

  \item \textbf{Extensive Evaluation.}
        We conduct detailed experiments and comparisons demonstrating that
        TrojanLoC preserves RTL semantics more effectively than graph-based
        methods and delivers significantly finer detection granularity,
        achieving higher precision, recall, and F1-score at both module and
        line levels.
\end{enumerate}

\section{Preliminaries} 
\label{sec:pre}
\subsection{Hardware Trojans}
\label{subsec:hardware-Trojans}

{HT}s are malicious modifications intentionally inserted into integrated circuits.
They are typically composed of two parts: a \emph{trigger} and a \emph{payload}.
The trigger is designed to activate only under rare internal states or input conditions so that the Trojan is hard to be detected during normal functional verification and production testing.
Once the trigger condition is met, the payload alters the circuit’s behavior or properties.
In practice, both the trigger and payload are often inserted into low-activity, small modules of the RTL, where signal toggling is infrequent and coverage by simulation and testing is weaker.
Trojan insertion can occur at different abstraction levels, including specification, RTL, gate-level netlists, or even during physical design and manufacturing, which makes early-stage detection particularly important.

HTs can be classified into four types according to their impacts~\cite{Salmani22}, i.e., functionality modification (T$1$), information leakage (T$2$), denial of service (T$3$), and performance degradation (T$4$).
This effect-based view highlights the diverse goals of HTs and
motivates the need for detectors that can not only identify the
presence of a Trojan but also distinguish its impact on the system.

\subsection{Decoder-only Transformer-based LLMs}
\label{subsec:transformer-prelim}
\begin{figure}[!bp]
    \centering
    \vspace{-6mm}
    \includegraphics[width=0.80\linewidth]{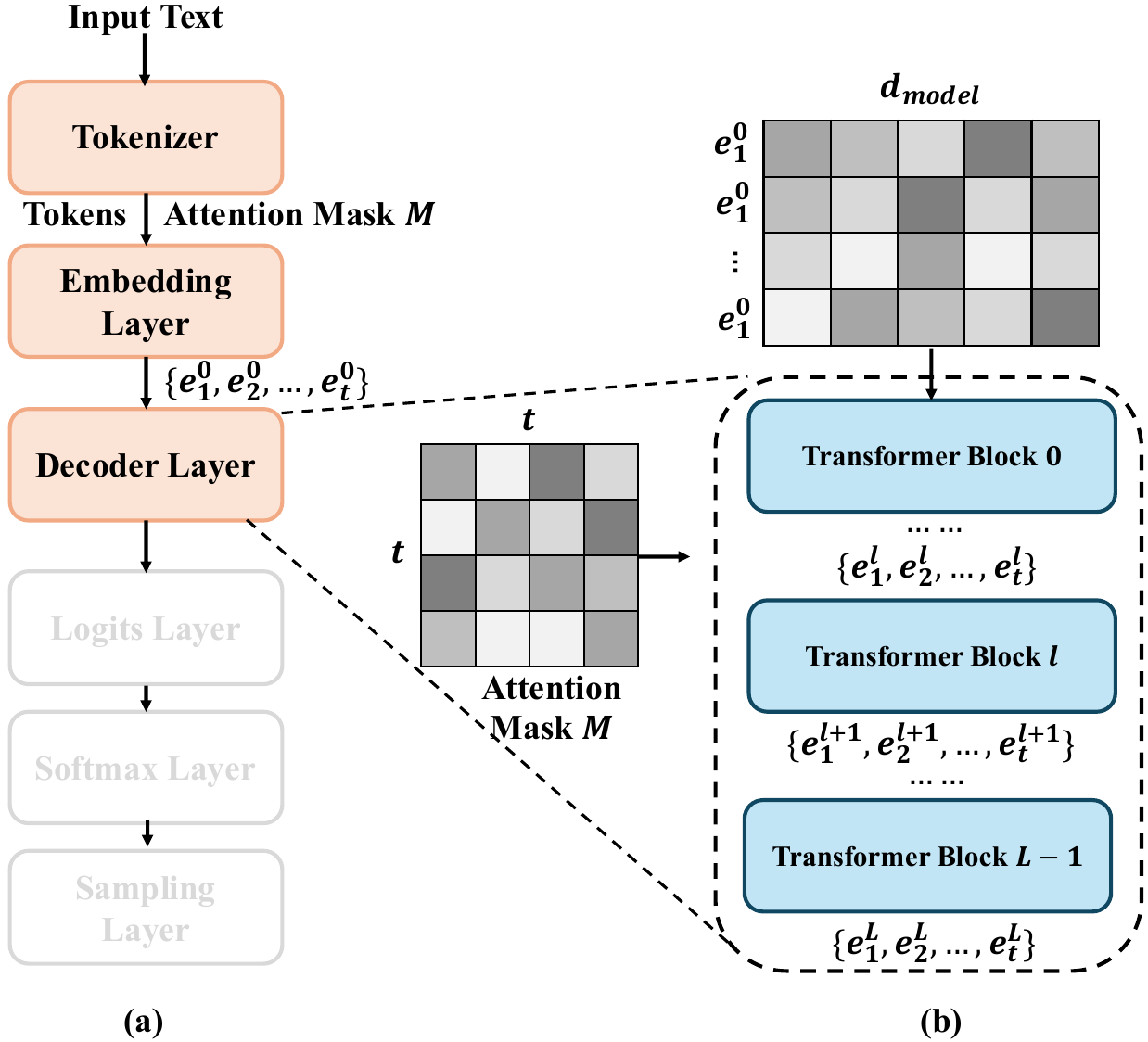}
    \caption{A decoder-only transformer-based LLM.}
    \label{fig:llm-architecture}
\end{figure}

\textit{Decoder-only transformer-based LLM}s have become the foundational architecture for modern generative and analytic AI systems.
In Fig.~\ref{fig:llm-architecture}(a), it shows the basic architecture of decoder-only transformer-based LLMs.
Given an input text, a \emph{tokenizer} first converts it into a sequence of $t$
discrete tokens, where $t$ is the sequence length.
An \textit{embedding layer} then maps each token to an embedding
$\mathbf{e}^{(0)}_k \in \mathbb{R}^{d_{\text{model}}}$ for $k = 1,\dots,t$, where
$d_{\text{model}}$ is a parameter of an LLM.
The input to the \textit{decoder layer} can therefore be viewed as a vector of embeddings
$\{\mathbf{e}^{(0)}_1,\dots,\mathbf{e}^{(0)}_t\}$ as shown in Fig.~\ref{fig:llm-architecture}.
The decoder layer consists of $L$ identical \textit{transformer block}s, which process these
embeddings in sequence (Fig.~\ref{fig:llm-architecture} (b)).

In transformer block $\ell$ (for $\ell = 0,\dots,L-1$), the current embeddings
$\{\mathbf{e}^{(\ell)}_1,\dots,\mathbf{e}^{(\ell)}_t\}$ are updated to
$\{\mathbf{e}^{(\ell+1)}_1,\dots,\mathbf{e}^{(\ell+1)}_t\}$.
For each position $k$, the block uses an attention module to mix information
from positions $j$, and then applies a small \textit{feed-forward network}. Which positions may interact is specified by an \emph{attention mask},
\[
M \in \{0,1\}^{t \times t},
\]
generated by the tokenizer as shown in Fig.~\ref{fig:llm-architecture}.
Each row $M_{k,*}$ is a length-$t$ vector that describes which tokens position $k$
is allowed to use: if $M_{k,j} = 1$ then token $j$ can contribute to the updated
embedding at position $k$, and if $M_{k,j} = 0$ it cannot.
In the causal setting, $M_{k,j} = 0$ for all $j > k$, so position $k$ only
uses information from tokens $1,\dots,k$.
When several input texts are concatenated and processed together, entries of $M$
are set to $0$ whenever tokens $k$ and $j$ belong to different texts.
In this way, one can think of $M$ as $t$ vectors (one per position), and the model
can process multiple texts in a single forward pass while keeping them independent.

After the last (i.e., $L$-th) transformer block, final embeddings
\[
\{\mathbf{e}^{(L)}_1,\dots,\mathbf{e}^{(L)}_t\},
\] are obtained
as one embedding $\mathbf{e}^{(L)}_k \in \mathbb{R}^{d_{\text{model}}}$ for each
token position $k$.
Each $\mathbf{e}^{(L)}_k$ can be viewed as the final embedding of token $k$,
because it encodes that token together with the context allowed by the mask.
In a standard LLM, these embeddings are passed through a layer for linear projection to \textit{vocabulary logit}s $\rightarrow$ a \textit{softmax layer} that converts
logits to probabilities over the next token $\rightarrow$a \textit{sampling layer} that selects the next token. We use the final token
embeddings $\{\mathbf{e}^{(L)}_k\}$ as the basis for higher-level
embeddings in Section~\ref{sec:methodology}.

\section{Methodology} \label{sec:methodology} 
The \textit{\textbf{TrojanLoC}} framework has three components:
\begin{enumerate}[leftmargin=*]
    \item [(1)] \textit{\textbf{TrojanInS dataset}} (Section~\ref{subsec:TrojanIns-Dataset}), which constructs a large-scale RTL corpus with systematically injected Trojans from four categories and fine-grained ground-truth annotations;
    \item [(2)] \textit{\textbf{LLM-based embedding extraction}} (Section~\ref{subsec:llm-embeddings}), which uses an RTL-finetuned LLM to derive module-level and line-level embeddings directly from TrojanInS RTL designs;
    \item [(3)] \textit{\textbf{Classifier training}} (Section~\ref{subsec:classifier-training}), which first trains an autoencoder to project these embeddings into a latent space to improve computational efficiency and robustness, and then trains task-specific classifiers on pairs of dimension-reduced embeddings and their corresponding labels for module-level Trojan detection/type prediction, and line-level Trojan localization.
     
\end{enumerate}

\subsection{TrojanInS Dataset}
\label{subsec:TrojanIns-Dataset}
Reliable Trojan detection needs a large RTL corpus with systematic insertions and fine-grained labels, yet existing benchmarks remain small and lack line-level detail. We introduce \textit{TrojanInS}, a dataset of $16000$+ Trojaned Verilog designs with line-level ground truth, whose statistics are shown in Table~\ref{tab:Trojanins-stats}.

\subsubsection{Base Dataset}
TrojanInS is constructed from VeriGen~\cite{Thakur24}, a collection of over $4000$ functionally correct Verilog designs gathered from public codebases and standard instructional materials. To ensure uniformity and avoid ambiguity in annotation, we restrict the dataset to designs containing a single top-level module, yielding roughly $4000$ clean circuits that span combinational logic, sequential control, and mixed-structure designs. This filtering guarantees consistent evaluation granularity and simplifies the mapping between Trojaned regions and the original design structure.

\subsubsection{Automated Trojan Insertion}
Trojan variants are generated through an LLM-guided insertion process in which GPT-4.1 introduces malicious behavior based on established hardware security taxonomies while keeping the design’s normal functionality unchanged. 
Each clean RTL file is expanded into four Trojaned versions, one for each category described in Section~\ref{subsec:hardware-Trojans}. 
These categories cover the major RTL-level Trojans studied in prior work and capture the core malicious behaviors relevant to detection and localization. For each design, the LLM receives the original Verilog code and an explicit Trojan specification, ensuring that the injected variants exhibit clear semantic and structural malicious patterns.

\subsubsection{Line-Level Annotation}
Each Trojan-inserted design in TrojanInS includes precise line-level labels that mark trigger logic, payload behavior, and any auxiliary modifications, while all unchanged lines are labeled as clean. We first identify newly inserted or altered regions in the LLM's output Trojaned design, then align them with the clean reference design to ensure accurate boundary marking. For every sample, the dataset provides (i) the clean RTL file, (ii) the Trojaned variant, (iii) a binary line-level mask, and (iv) metadata describing the Trojan type and its operational intent. 
These annotations allow us to train classifiers for multiple
tasks within a unified framework, including module-level Trojan detection,
Trojan-type classification, and fine-grained line-level localization, as shown in Section~\ref{sec:methodology}.

\begin{table}[t]
\centering
\caption{Statistics of the TrojanInS dataset.}
\label{tab:Trojanins-stats}
\small
\begin{tabular}{l c}
\toprule
\textbf{Property} & \textbf{Value} \\
\midrule
Base clean designs & $\sim 4{,}000$ \\
Trojaned designs & $> 17{,}000$ \\
Modules (train / test) & $17{,}658 \,/\, 4{,}407$ \\
Lines (train / test) & $1{,}964{,}853 \,/\, 496{,}347$ \\
Class distribution & Clean: $\approx 20\%$; \\
& Trojan: $\approx 80\%$ (T1--T4 balanced) \\
\bottomrule
\vspace{-6mm}
\end{tabular}
\end{table}

\subsection{LLM-based Embedding Extraction}
\label{subsec:llm-embeddings}

\begin{figure}[bp]
    \centering
    \includegraphics[width=\linewidth]{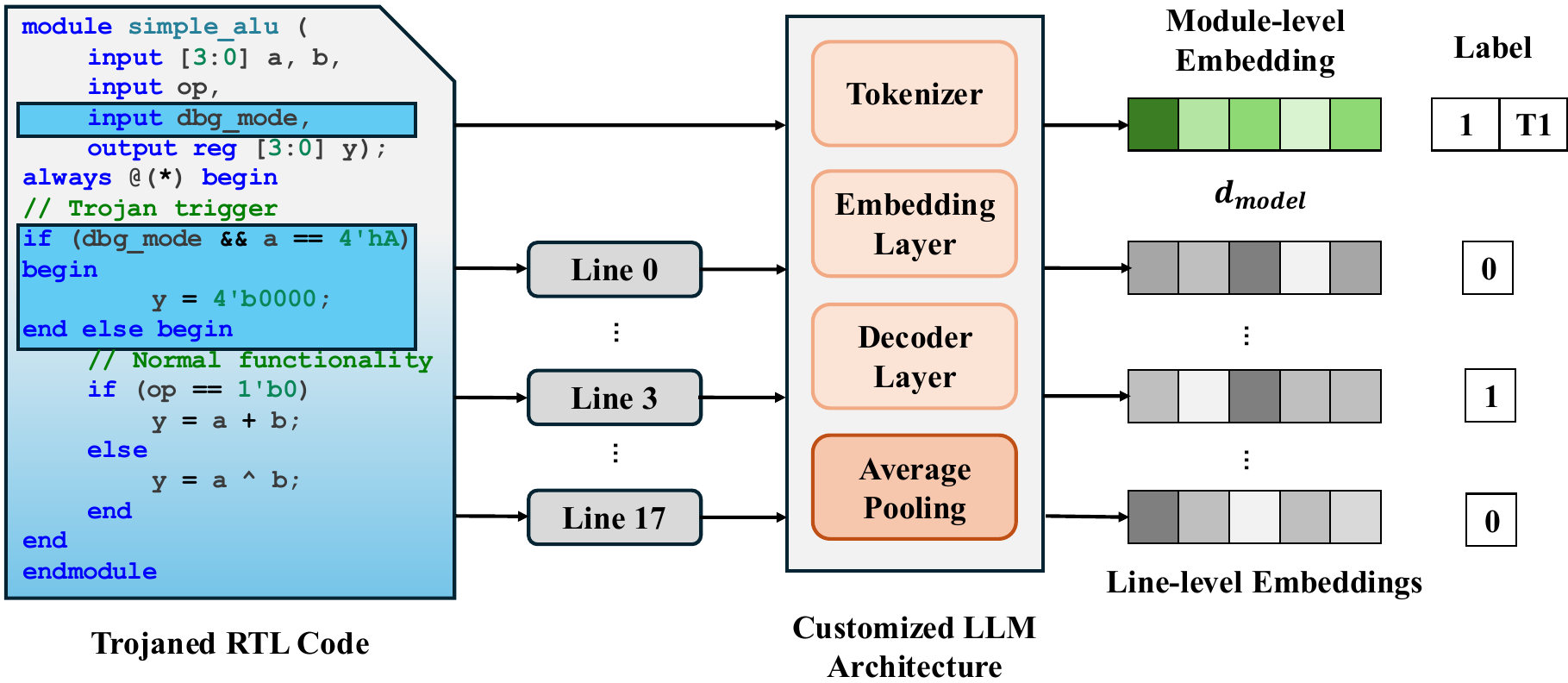}
    \caption{Overall flow of LLM-based embedding extraction.}
    \label{fig:llm-embedding-overview}
\end{figure}

TrojanLoC uses a decoder-only transformer LLM as an embedding extractor for RTL code.
We use the LLM that has been fine-tuned on a large amount of RTL codes, e.g., \textit{Verilog} and \textit{SystemVerilog}.
This RTL fine-tuning exposes the model to hardware-specific syntax, coding styles,
and common structural patterns (such as always blocks, reset logic, state machines,
and data-path operations).
As a result, the final token embeddings $\mathbf{e}^{(L)}_k$ can capture richer
semantics for hardware code than embeddings from a general LLM.
Figure~\ref{fig:llm-embedding-overview} illustrates the overall flow for processing an RTL module containing a T1 Trojan, with the Trojan-related lines highlighted by blue boxes.
A Trojaned RTL module (the same procedure is applied to clean modules) is first
split into its source lines.
Both the full module text and each individual line are then fed into a
\emph{LLM architecture} consisting of a tokenizer, an embedding
layer, a decoder layer with multiple transformer blocks, and an \textit{average pooling}
stage.
The LLM produces a single $d_{\text{model}}$-dimensional
\emph{module-level embedding} and a set of $d_{\text{model}}$-dimensional
\emph{line-level embeddings}, which are paired with module-level
and line-level labels and used as inputs to the autoencoders and classifiers in
Section~\ref{subsec:classifier-training}.

\subsubsection{Module-level Embeddings}
\label{subsubsec:module-embeddings}

For the module-level branch in Figure~\ref{fig:llm-embedding-overview}, the
entire RTL code of a module is treated as one input text.
The tokenizer converts this text into a sequence of $t$ tokens, and the
embedding layer maps them to initial embeddings
$\{\mathbf{e}^{(0)}_1,\dots,\mathbf{e}^{(0)}_t\}$.
These embeddings are processed by the decoder layer described in
Section~\ref{subsec:transformer-prelim}, which consists of $L$ transformer
blocks and outputs final token embeddings
$\{\mathbf{e}^{(L)}_1,\dots,\mathbf{e}^{(L)}_t\}$.

We then apply average pooling over all $t$ final token embeddings to obtain a
single vector that summarizes the module:
\[
\mathbf{z}_{\text{mod}} = \frac{1}{t} \sum_{k=1}^{t} \mathbf{e}^{(L)}_k
\in \mathbb{R}^{d_{\text{model}}}.
\]
This $\mathbf{z}_{\text{mod}}$ is shown as the green bar in
Figure~\ref{fig:llm-embedding-overview}.
In the figure, the module-level embedding is associated with a binary
label of $1$ and a type label T$1$, indicating that the module contains a T$1$ Trojan.

\subsubsection{Line-level Embeddings}
\label{subsubsec:line-embeddings}

For line-level localization, we extract one embedding per RTL line, as shown in
the lower branch of Figure~\ref{fig:llm-embedding-overview}.
Given a module with $L_{\text{line}}$ lines
$\ell_0,\ell_1,\dots,\ell_{L_{\text{line}}-1}$, each line $\ell_i$ is treated
as a text input.
The tokenizer converts $\ell_i$ into $t_i$ tokens, the embedding layer produces embeddings
$\{\mathbf{e}^{(0)}_{i,1},\dots,\mathbf{e}^{(0)}_{i,t_i}\}$, and the decoder
layer produces final embeddings
$\{\mathbf{e}^{(L)}_{i,1},\dots,\mathbf{e}^{(L)}_{i,t_i}\}$ for line. 

The line-level embedding is obtained by averaging its tokens:
\[
\mathbf{z}_{\text{line},i}
= \frac{1}{t_i} \sum_{k=1}^{t_i} \mathbf{e}^{(L)}_{i,k}
\in \mathbb{R}^{d_{\text{model}}}.
\]
In Figure~\ref{fig:llm-embedding-overview}, these are shown as the gray bars.
For efficiency, our implementation does not run the LLM separately for each
line.
Instead, multiple lines from a module are concatenated into a longer token
sequence and processed in a single forward pass, while an attention mask is
used to prevent tokens from different lines from attending to one another (as
described in Section~\ref{subsec:transformer-prelim}).
Conceptually, however, this procedure is equivalent to encoding each line
independently and then applying average pooling over its tokens.

Finally, each line embedding
$\mathbf{z}_{\text{line},i}$ is paired with a binary label
$y_{\text{line},i} \in \{0,1\}$ indicating whether line $\ell_i$ belongs to
Trojan logic.
In the example in Figure~\ref{fig:llm-embedding-overview}, line embeddings are
shown with labels $0$ or $1$ on the right, corresponding to clean and
Trojan-related lines, respectively.
These labeled embeddings are used by the line-level autoencoder and localization
classifier in Section~\ref{subsec:classifier-training}.

\subsection{Classifier Training}
\label{subsec:classifier-training}
\begin{figure}
    \centering
    \includegraphics[width=\linewidth]{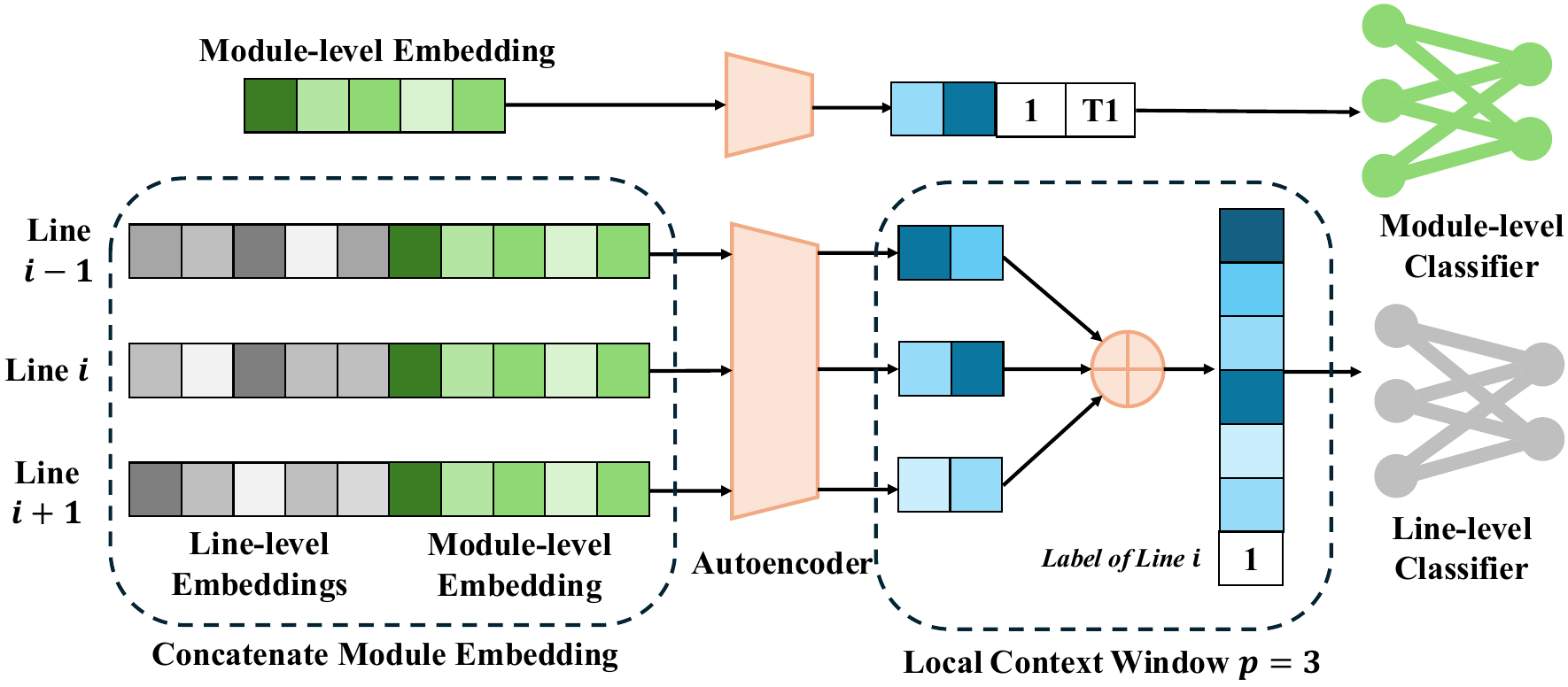}
    \caption{Training module- and line-level classifiers.}
    \label{fig:classifier-training}
    \vspace{-6mm}
\end{figure}
Figure~\ref{fig:classifier-training} gives the classifier
training stage in TrojanLoC.
The module-level embedding (top branch) and the line-level embeddings together
with the module embedding (bottom branch) are first compressed by a shared
autoencoder.
The resulting low-dimensional representations are then used by two separate
classifiers: a module-level classifier for Trojan detection and type
prediction, and a line-level classifier for Trojan localization.

\subsubsection{Autoencoders for Dimensionality Reduction}

To reduce the dimensionality of LLM embeddings and remove redundancy, TrojanLoC
uses two small autoencoders with the same architecture but different training
data: module and line-level embeddings.
The \emph{module-level} autoencoder takes the module embedding
\[
\mathbf{x}_{\text{mod}} = \mathbf{z}_{\text{mod}}
\in \mathbb{R}^{d_{\text{model}}}.
\]
For the \emph{line-level} autoencoder, each input combines local and global
information by concatenating the line embedding with the corresponding module
embedding:
\[
\mathbf{x}_{\text{line},i}
= [\,\mathbf{z}_{\text{line},i} \,;\, \mathbf{z}_{\text{mod}}\,]
\in \mathbb{R}^{2 d_{\text{model}}},
\]
as shown in the left dashed box of Figure~\ref{fig:classifier-training}.

In both cases, an encoder $f_{\text{enc}}(\cdot)$ maps the high-dimensional
input $\mathbf{x}$ to a lower-dimensional latent vector $\mathbf{h}\in \mathbb{R}^{d_{\text{enc}}}$ ($d_{\text{enc}} < d_{\text{model}}$), and a
decoder $f_{\text{dec}}(\cdot)$ reconstructs $\widehat{\mathbf{x}}$ from
$\mathbf{h}$.
Each autoencoder is trained with a standard reconstruction loss
\[
\mathcal{L}_{\text{AE}} = \bigl\| \mathbf{x} - \widehat{\mathbf{x}} \bigr\|_2^2,
\]
minimized over all modules (for $\mathbf{x}_{\text{mod}}$) or all lines (for
$\mathbf{x}_{\text{line},i}$).
Both autoencoders are trained only on the training split of TrojanInS. 
No test modules or lines are used during autoencoder training.
We include embeddings from both clean and Trojaned designs, since the autoencoders
serve purely for dimensionality reduction rather than Trojan detection.
After training, we discard the decoders and keep only the encoders, obtaining
compact features
\[
\mathbf{h}_{\text{mod}} = f_{\text{enc}}^{\text{mod}}(\mathbf{x}_{\text{mod}}),
\qquad
\mathbf{h}_{\text{line},i}
= f_{\text{enc}}^{\text{line}}(\mathbf{x}_{\text{line},i}),
\]
which correspond to the compressed vectors in the upper and lower branches of
Figure~\ref{fig:classifier-training}.
These reduced features are then used by the module-level and line-level
classifiers described next.

\subsubsection{Module-level and Line-level Classifiers}

\paragraph*{Module-level classifiers.}
Each module-level embedding $\mathbf{h}_{\text{mod}}$ is associated with two
labels: (i) a binary label indicating whether the module is clean or Trojaned
($0/1$), and (ii) a categorical label specifying the Trojan type (T1–T4) when a
Trojan is present.
For example, in Figure~\ref{fig:classifier-training}, the module embedding is
paired with the labels $1$ and T$1$, meaning that the module is Trojaned and the inserted Trojan is type T$1$.
To solve the two module-level tasks, we train two classifiers on $\mathbf{h}_{\text{mod}}$: one for Trojan detection (clean vs.\
Trojaned) and one for Trojan-type prediction (T1--T4).
Both classifiers use the corresponding module-level labels during training.

\paragraph*{Line-level classifier with local context.}
For line-level Trojan localization, we want to decide for each line~$i$ whether
it is part of Trojan logic.
Using only $\mathbf{h}_{\text{line},i}$ may miss important local patterns, so
we augment it with a context window of neighboring lines, as shown in the lower
branch of Figure~\ref{fig:classifier-training}.

Let $p$ denote the context window size.
For a target line index $i$, we take the reduced embeddings of lines
$i-\frac{p-1}{2},\dots,i,\dots,i+\frac{p-1}{2}$ and concatenate them to form a
context-augmented embedding:
\[
\mathbf{h}^{\text{aug}}_{\text{line},i}
  = [\,\mathbf{h}_{\text{line},i-\frac{p-1}{2}} \,;\, \dots \,;\,
       \mathbf{h}_{\text{line},i} \,;\, \dots \,;\,
       \mathbf{h}_{\text{line},i+\frac{p-1}{2}}\,].
\]
In Figure~\ref{fig:classifier-training}, this is illustrated with a window
$p=3$.
A line-level classifier takes
$\mathbf{h}^{\text{aug}}_{\text{line},i}$ as input and predicts whether line~$i$ belongs to Trojan logic.
During training of the line-level classifier, we use pairs of $\mathbf{h}^{\text{aug}}_{\text{line},i}$ and the corresponding binary label $y_{\text{line},i} \in \{0,1\}$.

Overall, the pipeline in Figure~\ref{fig:classifier-training} shows how a single
autoencoder provides compact representations for both module- and line-level
embeddings, and how these embeddings, combined with module-level information and
local line context, are used to train classifiers to perform Trojan
detection, Trojan-type prediction, and fine-grained line-level Trojan localization.

\section{Experimental Evaluation}
\label{sec:experiments}

In this section, we evaluate TrojanLoC on both module-level and line-level
HT tasks.
We first describe the experimental setup, including datasets, embedding
backbones, and classifiers.
We report results for module-level Trojan detection and Trojan-type
prediction, followed by a study of line-level Trojan localization under
different design choices and hyperparameters.

\subsection{Experimental Setup}
\label{subsec:exp-setup}

\textbf{Dataset.}
All experiments are conducted on the RTL dataset described in
Section~\ref{sec:methodology}.
We split the dataset into $80\%$ for training and $20\%$ for testing. Module-level tasks (Trojan detection and type prediction) are evaluated on
modules, while line-level localization is evaluated on RTL lines with
binary labels. To ensure our model learns semantic patterns rather than superficial cues, we preprocess the RTL code by removing Trojan-related comments and replacing Trojan-indicative variable names with benign alternatives. 
This step is implemented by a Python script and does not use an LLM.
This prevents the model from solving tasks through keyword matching.

\textbf{Embedding backbones.}
Unless otherwise specified, TrojanLoC uses decoder-only transformer LLMs that
have been fine-tuned on large RTL corpora (e.g., Verilog and
SystemVerilog) to produce module-level and line-level embeddings, as described
in Section~\ref{subsec:llm-embeddings}.
In experiments, we evaluate TrojanLoC using three finetuned LLMs: \textit{CL-Verilog} 13B~\cite{Nakkab24}, \textit{CodeV-QW} 7B~\cite{zhao24}, \textit{HaVen-CodeQWen} 7B~\cite{Yang25}.

\textbf{Dimensionality reduction.}
For all settings, we use the autoencoders described in
Section~\ref{subsec:classifier-training} to reduce the dimensionality of
module-level embeddings $\mathbf{z}_{\text{mod}}$ and combined line-level
embeddings $\mathbf{x}_{\text{line},i}$.
The dimensions of $\mathbf{h}_{\text{mod}}$ and $\mathbf{h}_{\text{line,i}}$ are both $128$.

\textbf{Classifiers.}
We employ tree-based gradient boosting models, XGBoost~\cite{Chen16} and LightGBM~\cite{Ke17}, as final classifiers due to their robustness and ability to handle mixed and redundant features. We use XGBoost~\cite{Chen16} and LightGBM~\cite{Ke17} for all three tasks:
(i) binary module-level Trojan detection,
(ii) multi-class module-level Trojan-type prediction, and
(iii) binary line-level Trojan localization.

\textbf{Baselines.}
For module-level tasks, we compare TrojanLoC against:
(i) GNN4TJ~\cite{Yasaei2021}, with a GNN to extract embeddings from the DFG of an RTL design,
(ii) TrojanSAINT~\cite{Lashen23}, with a GNN to extract embeddings from the gate-level netlist of an RTL design. 
We derive the gate-level netlist by synthesizing an RTL design with the Synopsys 90nm Generic Library through Yosys~\cite{wolf2013yosys}\footnote{We use the Synopsys generic library created for educational purposes instead of a foundry-provided standard cell library, as using the latter may result in NDA violation.}, and
(iii) LLM-based Trojan detection baseline~\cite{Hayashi25}, where a general LLM is prompted
with the RTL module and asked whether the module is Trojaned.

\textbf{Metrics.}
There are four primary evaluation metrics used in the experimental section: accuracy (denote as Acc), precision (denoted as P), recall (denoted as R), and $\text{F}1$.
Given \textit{true positives} (\textit{TP}), \textit{false positives} (\textit{FP}), and \textit{false negatives} (\textit{FN}),
they are defined as:
\[
\text{P} = \frac{\mathrm{TP}}{\mathrm{TP} + \mathrm{FP}}, \qquad
\text{R} = \frac{\mathrm{TP}}{\mathrm{TP} + \mathrm{FN}},
\]
\[
\text{F}1_{\text{clean}} = \frac{2 \cdot \text{P} \cdot \text{R}}
{\text{P} + \text{R}}, \qquad
\text{Acc} = \frac{\mathrm{TP} + \mathrm{TN}}
{\mathrm{TP} + \mathrm{FP} + \mathrm{TN} + \mathrm{FN}}.
\]

For module-level Trojan detection, we report precision, recall, and F$1$ for the Trojan class.
For module-level Trojan-type prediction, we report accuracy, precision, recall, and F$1$ over the four Trojan categories (T$1$--T$4$).
For line-level Trojan localization, we treat each line as a sample and report F$1$ for the non-Trojan and Trojan classes.

\textbf{Hardware.}
All embedding extraction and classifier training in our experiments are
performed on a single NVIDIA H$100$ GPU.

\subsection{Module-Level Trojan Detection}
\label{subsec:exp-module}

We evaluate TrojanLoC on two module-level tasks: (i) binary Trojan
detection (clean vs.\ Trojan), and (ii) Trojan-type prediction (T1--T4).
Unless noted, 
all module embeddings $\mathbf{z}_{\text{mod}}$ are
projected through the module autoencoder to $\mathbf{h}_{\text{mod}}$ and
used as classifier inputs.

\subsubsection{Comparison with Graph and Prompting Baselines}
\label{subsec:rq1-baselines}

TrojanLoC yields strong improvements across all metrics. Table~\ref{tab:module-detection} compares TrojanLoC against graph-based baselines (GNN4TJ, TrojanSAINT) and prompting baselines using GPT-4o and GPT-4o-mini. 
Graph-based methods fail due to inherent limitations. GNN4TJ cannot process 14.09\% of designs due to its DFG transformations and lack of submodule definitions; even for processed designs, it achieves only 0.27 accuracy, as DFG embeddings cannot capture Trojan patterns. TrojanSAINT performs node-level classification, and we mark a design as Trojaned if any node is predicted malicious. However, 21.24\% of designs fail to synthesize because they are non-top-level modules lacking complete module hierarchy information required for synthesis. 
For processed designs, it achieves $0.80$ accuracy.

We evaluate the direct LLM-based Trojan detection baseline over GPT-4o-mini and GPT-4o.
Both GPT-4o-mini and GPT-4o achieve relative high accuracy $0.87$ and $0.94$ respectively, and high $\text{F}1_\text{Trojan}$ $0.91$ and $0.96$ respectively.
However, for the prediction of the clean class, they achieve a relative low performance with $\text{F}1_\text{clean}$ $0.75$ and $0.87$ respectively, which indicates that generic code reasoning of general LLMs is insufficient for capturing subtle RTL trigger/payload patterns.
In contrast, all RTL-finetuned variants of TrojanLoC achieve $\text{F}1_\text{Trojan}=0.98\text{--}0.99$ and $\text{F}1_\text{clean}=0.93\text{--}0.96$, showing that domain-adapted LLM embeddings provide the semantic granularity required for robust Trojan detection.

\begin{table}[t]
\centering
\caption{Module-level HT detection: baselines vs.\ TrojanLoC.}
\label{tab:module-detection}
\small
\begin{tabular}{lcccccc}
\toprule
\textbf{Model} &
\textbf{Acc} &
\textbf{$\text{P}_{\text{Trojan}}$} &
\textbf{$\text{R}_{\text{Trojan}}$} &
\textbf{$\text{F}1_{\text{Trojan}}$} &
\textbf{$\text{F}1_{\text{clean}}$}\\
\midrule
\multicolumn{6}{l}{\textit{\textbf{Baseline}}}\\[-2pt]
GNN4TJ & 0.27 & 0.66 & 0.20 & 0.31 &  0.23 \\
TrojanSAINT & 0.80 & 0.84 & 0.91 & 0.88 & 0.47  \\
\tabincell{c}{GPT-4o-mini} & 0.87 & 0.98 & 0.85 & 0.91 & 0.75 \\
\tabincell{c}{GPT-4o} & 0.94 & 0.99 & 0.94 & 0.96 & 0.87 \\
\midrule
\multicolumn{6}{l}{\textit{\textbf{TrojanLoc}}}\\[-1pt]
CLVerilog-XGB &
0.98 & 0.99 & 0.99 & \textbf{0.99} & \textbf{0.96}  \\
CLVerilog-LGBM &
0.99 & 0.98 & 0.99 & \textbf{0.99} & \textbf{0.96} \\
CodeV-QW-XGB &
0.97 & 0.98 & 0.99 & \textbf{0.98} & \textbf{0.93}  \\
CodeV-QW-LGBM &
0.97 & 0.98 & 0.99 & \textbf{0.98} & \textbf{0.93} \\
\tabincell{c}{HaVen-CodeQWen\\-XGB} &
0.98 & 0.99 & 0.99 & \textbf{0.99} & \textbf{0.95}  \\
\tabincell{c}{HaVen-CodeQWen\\-LGBM} &
0.98 & 0.99 & 1.00 & \textbf{0.99} & \textbf{0.96} \\
\bottomrule
\vspace{-7mm}
\end{tabular}
\end{table}


\subsubsection{Effect of RTL-Finetuned LLMs and Classifiers}
\label{subsec:rq1-ablation}

Table~\ref{tab:module-detection} presents an ablation across LLM
backbones and gradient-boosted classifiers. RTL-finetuned LLMs
(CLVerilog, CodeV-QW, HaVen-CodeQWen) consistently outperform direct prompting by a wide margin, confirming that exposure to hardware syntax, common coding patterns is crucial for learning Trojan-related semantics. 
Differences between the three
RTL-finetuned LLMs are minor, suggesting that our TrojanLoC can be applicable to different LLMs finetuned using different training datasets.
Classifier choice minimally affects binary detection
($\text{F}1_\text{Trojan}\approx0.99$ across all settings). 






\subsection{Module-Level Trojan Type Prediction}
\label{subsec:exp-type}

Beyond binary detection, TrojanLoC aims to classify the 
effect-based Trojan category (T1--T4) of the module. This task is more challenging, as it entails distinguishing subtle semantic differences in the trigger
and payload logic that define a Trojan family.

\subsubsection{Type Classification Performance}
\label{subsec:rq1-type-overall}
Table~\ref{tab:module-type} summarizes TrojanLoC's performance across all backbones and classifiers. Accuracy and macro-F1 range from $0.80$ to
$0.84$, with best results from LightGBM paired with CodeV-QW-7B or HaVen-CodeQWen-7B. The embeddings retain sufficient semantic resolution to separate the four categories, despite many Trojans modifying only a few RTL lines. Notably, the narrow performance range indicates minimal impact from model size, fine-tuning effectively captures normal design patterns, enabling strong clean/Trojan distinction (Table~\ref{tab:module-detection}) while maintaining consistent type classification across model scales.

We observe moderate performance variation across backbones. Models with stronger RTL specialization yield more discriminative embeddings for type classification. This trend reflects each model's training emphasis, HaVen-CodeQWen-7B prioritizes functional consistency and structural reasoning, directly benefiting anomaly-oriented pattern recognition such as HT types; CodeV-QW emphasizes instruction fidelity; while CL-Verilog-13B is largely syntax-driven, resulting in progressively weaker structural discrimination. Classifier choice has a secondary effect. LightGBM consistently outperforms XGBoost in macro-F1, suggesting its leaf-wise splitting captures finer boundaries between Trojan types.

\begin{figure}[htbp]
    \centering
    \includegraphics[width=0.7\linewidth]{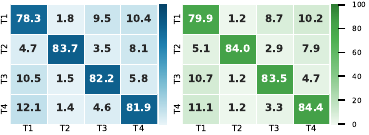}
    \caption{Average accuracy for XGB (L) and LGBM (R) on CLVerilog, CodeQWen and CodeV for module-level detection.}
    \label{fig:class_heatmap}
\vspace{-6mm}
\end{figure}

\subsubsection{Per-Type Analysis and Confusion Patterns}
\label{subsec:rq1-type-confusion}

A per-type analysis reveals performance patterns. Information leakage Trojans (T2) and performance degradation Trojans (T4) achieve the highest accuracies, $\approx$84\%, driven by their distinctive routing and timing signatures. Functional-modification Trojans (T1) remain the most challenging, with accuracies $\approx 80\%$, while denial of service Trojans (T3) fall in range near 83\%. As shown in Figure~\ref{fig:class_heatmap}, both classifiers display comparable diagonal accuracy, though LightGBM offers sharper separation for T2 and T4. This advantage is evident for T4, where LightGBM reaches 84.4\% compared to 81.9\% for XGBoost, showing stronger sensitivity to timing and routing Trojans.

The confusion matrix highlights the structural proximity among categories. T1 is often mistaken for T3 and T4, consistent with their shared reliance on conditional control-flow changes that reshape datapath behavior. These overlaps blur semantic boundaries and make fine-grained discrimination difficult. In contrast, T2 exhibits the clear separation with minimal cross-type confusion, underscoring the distinctiveness of its leakage-oriented logic patterns.

\begin{table}[t]
\centering
\caption{Module-level HT type prediction (T1--T4). Reported: 
overall accuracy, macro-averaged precision/recall/F1.}
\label{tab:module-type}
\small
\begin{tabular}{lccccc}
\toprule
\textbf{Model} &
\textbf{Acc} &
\textbf{$\text{P}_{\text{macro}}$} &
\textbf{$\text{R}_{\text{macro}}$}  &
\textbf{$\text{F1}_{\text{macro}}$} \\
\midrule
CLVerilog-XGB &
0.80 & 0.81 & 0.80 & 0.80 \\
CLVerilog-LGBM &
\textbf{0.81 }& \textbf{0.82} & \textbf{0.81} & \textbf{0.81}  \\
\midrule
CodeV-QW-XGB &
0.82 & 0.83 & 0.82 & 0.82  \\
CodeV-QW-LGBM &
\textbf{0.84} & \textbf{0.84} & \textbf{0.84} & \textbf{0.84} \\
\midrule
HaVen-CodeQWen-XGB &
0.82 & 0.83 & 0.82 & 0.82 \\
HaVen-CodeQWen-LGBM &
\textbf{0.84} & \textbf{0.85} & \textbf{0.84} & \textbf{0.84} \\
\bottomrule
\end{tabular}
\end{table}


\subsection{Line-Level Trojan Localization}
\label{subsec:exp-line}

We now evaluate TrojanLoC on fine-grained Trojan localization,
where the goal is to identify the specific RTL lines responsible for
trigger or payload behavior. This task is considerably more
demanding than module-level prediction, as Trojan lines are sparse,
often resemble benign control or assignment statements, and may
differ from clean logic only in subtle conditional dependencies.

\subsubsection{Effect of Module Embedding and Context}
\label{subsec:rq2-ablation}
To understand design choices behind line-level performance, we perform controlled ablation using CLVerilog+XGBoost, varying (i) module-level embedding concatenation ($m\in\{0,1\}$) and (ii) symmetric context window size $p$, shown in Tab.~\ref{tab:line-ablation}.
\textbf{(1) Impact of context window size}:
Context is crucial. Increasing $p$ from $0\rightarrow3$ raises $\mathrm{F1}_{\text{Trojan}}$ from $0.81$ to $0.88$, saturating around $p=3\text{--}5$. This matches RTL coding practice: Trojans typically span short multi-line patterns (e.g., trigger condition followed by payload assignment) rarely extending beyond $\pm 5$ lines. Larger windows provide diminishing returns and introduce noise.
\textbf{(2) Impact of module embeddings}:
Without context ($p=0$), module embedding offers minimal benefit ($\mathrm{F1}_{\text{Trojan}}=0.81$ for both $m=0,1$). With context, module embeddings consistently help---at $p=5$, $\mathrm{F1}_{\text{macro}}$ rises from $0.92$ to $0.93$ when $m=1$. This shows global semantics become useful only when combined with local context: line embeddings capture local behavior while module embeddings provide global consistency cues.

\subsubsection{Localization Results}
\label{subsec:rq2-main}
Table~\ref{tab:line-ablation} shows the best configuration for each RTL-finetuned backbone. TrojanLoC achieves $\mathrm{F1}_{\text{clean}}=0.96$ and $\mathrm{F1}_{\text{Trojan}}=0.88\text{--}0.89$, yielding $\mathrm{F1}_{\text{macro}}=0.92\text{--}0.93$. These results confirm that RTL-tuned embeddings retain token-level semantics for line-level reasoning beyond coarse module classification. The $\sim$8-point gap reflects the difficulty of identifying minimally intrusive malicious lines differing from benign logic by only small conditionals or subtle assignments. TrojanLoC's suspiciousness scores enable analysts to review only the top few percent of ranked lines, true Trojan lines consistently appear within the top 3--5\%, reducing manual auditing by over an order of magnitude.

\begin{table}[t]
\centering
\caption{Ablation on line-level localization using CLVerilog + XGBoost, where 
$m$=module embedding, $p$=context window.}
\setlength{\tabcolsep}{2.5pt}
\label{tab:line-ablation}
\small
\begin{tabular}{@{}l@{\hspace{6pt}}l@{\hspace{8pt}}cc@{\hspace{12pt}}cc@{\hspace{12pt}}cc@{}}
\toprule
& & \multicolumn{2}{c}{$p$=0} & \multicolumn{2}{c}{$p$=3} & \multicolumn{2}{c}{$p$=5} \\
\cmidrule(lr){3-4} \cmidrule(lr){5-6} \cmidrule(l){7-8}
Model & Metric & $m$=0 & $m$=1 & $m$=0 & $m$=1 & $m$=0 & $m$=1 \\
\midrule
\multirow{3}{*}{CLVerilog-XGB}
  & \textbf{$F1_\text{clean}$}  & 0.94 & 0.93 & \textbf{0.96} & \textbf{0.96} & \textbf{0.96} & \textbf{0.96} \\
  & \textbf{$F1_\text{Trojan}$} & 0.81 & 0.81 & 0.87 & 0.88 & 0.88 & \textbf{0.89} \\
  & \textbf{$F1_\text{macro}$}  & 0.88 & 0.87 & 0.92 & 0.92 & 0.92 & \textbf{0.93} \\
\midrule
\multirow{3}{*}{CodeV-QW-XGB}
  & \textbf{$F1_\text{clean}$}  & 0.94 & 0.94 & \textbf{0.96} & \textbf{0.96} & \textbf{0.96} & \textbf{0.96} \\
  & \textbf{$F1_\text{Trojan}$} & 0.80 & 0.82 & \textbf{0.88} & \textbf{0.88} & 0.87 & \textbf{0.88} \\
  & \textbf{$F1_\text{macro}$}  & 0.87 & 0.88 & \textbf{0.92} & \textbf{0.92} & \textbf{0.92} & \textbf{0.92} \\
\midrule
\multirow{3}{*}{\shortstack{HaVen-CodeQWen-\\XGB}}
  & \textbf{$F1_\text{clean}$}  & 0.94 & 0.93 & \textbf{0.96} & \textbf{0.96} & \textbf{0.96} & \textbf{0.96} \\
  & \textbf{$F1_\text{Trojan}$} & 0.81 & 0.80 & \textbf{0.89} & 0.88 & 0.88 & 0.88 \\
  & \textbf{$F1_\text{macro}$}  & 0.88 & 0.87 & \textbf{0.93} & 0.92 & 0.92 & 0.92 \\
\bottomrule
\end{tabular}
\vspace{-5mm}
\end{table}

\section{Conclusion} \label{sec:con}
We presented TrojanLoC, an LLM-based framework for fine-grained HT detection and localization directly from RTL code. Built on TrojanInS, a dataset with 17k+ designs covering four major Trojan families, TrojanLoC extracts module and line level embeddings using RTL-finetuned LLMs and employs lightweight classifiers for unified binary detection, type prediction, and line-level localization. By avoiding lossy netlist or graph conversions, TrojanLoC preserves RTL semantics and achieves 0.99 F$1$-score for module-level detection and 0.92 macro-F$1$ for line-level localization, outperforming graph-based baselines. Its ranked suspiciousness scores surface true Trojan lines within the top few percent, reducing manual auditing effort by over an order of magnitude and offering a practical solution for scalable hardware security validation.

\bibliographystyle{ACM-Reference-Format}
\bibliography{Reference}

\end{document}